# Assay methods for $^{238}$U, $^{232}$Th, and $^{210}$Pb in lead and calibration of $^{210}$Bi bremsstrahlung emission from lead

John L. Orrell[1], Craig E. Aalseth[1], Isaac J. Arnquist[1], Tere A. Eggemeyer[1],

Brian D. Glasgow[1], Eric W. Hoppe[1], Martin E. Keillor[1], Shannon M. Morley[1],

Allan W. Myers[1], Cory T. Overman[1], Sarah M. Shaff[1], and Kimbrelle S. Thommasson[1]

*[1]National Security Directorate, Pacific Northwest National Laboratory,*

*902 Battelle Boulevard, P.O. Box 999, MSIN J4-65, Richland, WA 99352 USA*

## Abstract

Assay methods for measuring $^{238}$U, $^{232}$Th, and $^{210}$Pb concentrations in refined lead are presented. The $^{238}$U and $^{232}$Th concentrations are determined using isotope dilution inductively coupled plasma mass spectrometry (ID-ICP-MS) after anion exchange column separation of dissolved lead samples. The $^{210}$Pb concentration is inferred through α-spectroscopy of a daughter isotope, $^{210}$Po, after chemical precipitation separation on dissolved lead samples. Subsequent to the $^{210}$Po α-spectroscopy assay, a method for evaluating $^{210}$Pb concentrations in solid lead samples was developed via measurement of bremsstrahlung radiation from β-decay of a daughter isotope, $^{210}$Bi, by employing a 14-crystal array of high purity germanium (HPGe) detectors. Ten sources of refined lead were assayed. The $^{238}$U concentrations were <34 microBq/kg and the $^{232}$Th concentrations ranged <0.6 – 15 microBq/kg, as determined by the ICP-MS assay method. The $^{210}$Pb concentrations ranged from ~0.1 – 75 Bq/kg, as inferred by the $^{210}$Po α-spectroscopy assay method.

## Keywords

- Lead assay

- $^{238}$U, $^{232}$Th, $^{210}$Pb

- Inductively coupled plasma mass spectrometry (ICP-MS)

- α-spectroscopy

- γ-ray spectroscopy

## Introduction

Refined lead is used as a γ-ray shielding material in many radiation measurement systems. In low background radiation measurement systems, selecting lead with the lowest radioactivity is an important consideration for achieving the highest sensitivity in the measurement system [1-4]. The radioactive impurities in lead are primarily $^{238}$U and $^{232}$Th decay-chain isotopes, which are present naturally in the ore from which lead is refined. The radioactive lead isotope, $^{210}$Pb, a daughter isotope in the $^{238}$U decay chain, is substantially out of secular equilibrium with the $^{238}$U concentration in post-refinement lead. This is a consequence of the elemental fractionation refinement used to process lead. Furthermore, the isotopic fraction of $^{210}$Pb present in refined lead cannot be reduced through subsequent chemical processing. For these reasons, assay of the residual radioactivity in refined lead should consider each isotope, $^{238}$U, $^{232}$Th, and $^{210}$Pb, as characteristic of the lead's utility as a low background shielding material for high sensitivity radiation measurement systems.

## Experimental

This section presents the two principal analytical chemistry techniques used for measurement of $^{238}$U, $^{232}$Th, and $^{210}$Pb in acid dissolutions of samples of refined lead. The determination of $^{238}$U and $^{232}$Th in refined lead utilizes anion exchange separation in order to pre-concentrate the analyte and substantially remove much of the lead matrix previous to analysis by ICP-MS. Isotope dilution methods were employed using $^{233}$U and $^{229}$Th tracer solutions for quantifying the assay results. This ICP-MS method follows the work of LaFerriere *et al.* [5]. The determination of $^{210}$Pb in refined lead was inferred from the α-particle emission of its $^{210}$Po daughter. A $^{209}$Po tracer was used to quantify the α-spectroscopy measurement. After the bulk of the lead matrix was removed via a precipitation reaction, the Po was isolated using an extraction resin separation and the remaining Po was electrochemically deposited onto a Ag disk. This α-spectroscopy method is a development from the prior work of Miley *et al.* [6]. It is worth noting this $^{210}$Po-based $\alpha$-spectroscopy method assumes secular equilibrium is established in the lead samples through both $^{210}$Bi (5 day half-life) and $^{210}$Po (138 day half-life). If a lead sample is taken from a recently refined source (within several months of refinement), it is advantageous to use the bremsstrahlung emission from $^{210}$Bi beta decay as a more accurate estimator of the $^{210}$Pb activity level. A method of using bremsstrahlung emission from $^{210}$Bi beta decay measured via $\gamma$-ray spectroscopy to estimate $^{210}$Pb activity levels is developed using the results from the $^{210}$Po-based $\alpha$-spectroscopy method to provide a set of calibration sources. In all cases described, the lead used for these assays was cut from the center of a brick or ingot (i.e., aliquot cubes with sides of approximately 1 cm length) so no surface contaminants would enter the assay process and therefore the results are representative of the "bulk" of the lead.

## Methodology for Determination of $^{238}$U and $^{232}$Th in Lead using ICP-MS

*Facilities.* All sample preparation for the analysis of $^{238}$U and $^{232}$Th was performed in a Class 10,000 cleanroom at Pacific Northwest National Laboratory (PNNL). Moreover, samples were prepared and separations performed in a Class 10 laminar flow hood.

*Chemicals and Reagents.* All chemicals and reagents used throughout the sample preparation for ICP-MS analysis were acquired at the highest purity. Deionized, triply distilled water was used exclusively.

*Instrumentation.* Analyses were performed using an Agilent 7700x series ICP-MS (Agilent Technologies, Santa Clara, CA, USA) with an integrated autosampler equipped with an Agilent microflow nebulizer and standard quartz double-pass spray chamber. Plasma, ion optics, and mass analyzer settings were tuned to maximize sensitivity for high m/z species (specifically U and Th) at the expense of the lower mass range. During tuning, oxides were monitored and typically limited to a $CeO^+/Ce^+$ of less than 1.5%. Daily tuning and optimization was performed as needed. In order to improve instrumental precision and limit the associated error due to counting statistics, data acquisition times were optimized for analyte and radiotracer. Instrumental acquisitions were acquired in triplicate for each sample.

*Leaching and Validation of Labware.*  All labware was cleaned and validated to instrument background levels for $^{238}$U and $^{232}$Th. The validation process was performed as follows: labware was soaked in 6M $HNO_3$ in a large sealable container for a minimum of 24 hours in a vacuum oven at 80°C. All labware was then triply rinsed in deionized distilled water. Labware containers were then filled with *ca.*1.5 mL of 5% (v/v) $HNO_3$ and submitted to an additional overnight 80°C leaching. Leachates were then analyzed for $^{238}$U and $^{232}$Th on the ICP-MS to ensure insignificant background contribution due to unclean labware. Labware passing validation was then rinsed three times using deionized distilled water and dried with filtered nitrogen. If labware failed validation, additional repetitions of the above cycle would be performed until the material passed validation before using.

*Sample Dissolution and Preparation.*  Approximately 200-400 mg portions of the lead samples were removed using a new and clean stainless steel razor blade. In order to remove any surface contamination, a surface etch of the sample was performed in 10% (v/v) $HNO_3$ in an oven at 80°C. After etching, the samplings were rinsed with deionized distilled water, dried with filtered nitrogen gas, and weighed. Full digestions of the samplings were performed in validated Savillex vials (Bloomington, MN, USA) with 10% (v/v) $HNO_3$ spiked with a known amount of radioactive $^{229}$Th and $^{233}$U tracer. After complete dissolution and in preparation for column loading, all sample solutions were diluted up to a final volume of 5 mL of the same nominal lead concentration in a 6 M $HNO_3$ matrix.

*Column and Resin Preparation.*  The following described column cleaning and separation procedure follows that developed in LaFerriere *et al* [5]. Briefly, columns of 500 μL bed volumes were prepared from AG 1×4, 100-200 anion exchange resin (Bio-Rad Corp., Hercules, CA, USA). Columns were purchased from Environmental Express (Charleston, SC, USA) and had the following dimensions: 7.1-mm i.d., 93.6-mm length, and a 2.5-mL bed capacity. All columns and included frits were leached at room temperature for at least 72 hours in 6 M $HNO_3$ and triply rinsed in deionized distilled water prior to use.

Prior to loading the leached columns with resin, the ion exchange resin was cleaned in 10 mL bulk batches as described previously [5]. Column validations for radiopurity were performed to ensure limited sample preparation contributions. Columns were not reused due to the difficulty of removing lead after a single use.

*Sample loading and separation and columns.* Resin conditioning was performed by passing 2 mL of 6 M $HNO_3$ through the column just prior to loading the sample. The prepared sample (in 6 M $HNO_3$ matrix) was then loaded onto the column. The resin was washed with 1.8 mL of 6 M $HNO_3$ in order to substantially clean the resin of bound lead species. Finally, the analytes were eluted and collected off the column using 1.2 mL of 0.1 M HCl and analyzed using ICP-MS.

## Methodology for Determination of $^{210}$Po in Lead using $\alpha$-spectroscopy

*Chemicals and Reagents.* All of the chemicals used during this experiment were Fisher Chemical Optima$^{\copyright}$ grade chemicals. The $^{209}$Po standard was procured from Eckert and Ziegler (Valencia, CA, USA) and diluted in 3 M $HNO_3$ such that its average activity was 1.0 pCi/mL, or 0.037 Bq/mL. All labware used during this experiment were triply rinsed with doubly distilled de-ionized water and air-dried. The dissolution vessels were leached in 1% (v/v) $HNO_3$ before each use to avoid carryover contamination between samples. Hot plate temperatures were monitored with a Spot-Check surface thermometer throughout the experiment to avoid volatilization of $PoNO_3$.

*Sample Preparation.* Four sample aliquots of approximately 10 g were machined from ten different sources of Pb bricks. Each was cleaned to remove any surface contamination and the outer oxide layer prior to dissolution. This was done by cleaning the lead with a small amount of 8 M $HNO_3$ and a few drops of 30% $H_2O_2$ in a beaker over a hot plate. After this reaction was complete, the lead was rinsed with deionized water, dried, and weighed. For each of the ten different sources of lead, a process blank and a reagent blank were also carried through the procedure with the four sample aliquots, resulting in a set of six for each lead sample.

*Sample Dissolution.* The bulk lead dissolution was completed in a CEM Microwave Accelerated Reaction System 5 (MARS5). Sample sets of six were dissolved at one time. The lead pieces were placed in Teflon microwave vessels and covered with 8 mL 8M $HNO_3$. A few drops of 30% $H_2O_2$ were added initially, and more acid and peroxide were added between runs in the microwave as needed until the lead was completely dissolved. All of the sample aliquots in a set were exposed to the same amount of time in the microwave. The microwave was set to ramp up to 150°C over 15 min with a maximum pressure of 5 atm. Then the instrument held the temperature at 150°C for 15 min with a maximum pressure of 5 atm.

*Precipitation.* Once the lead was dissolved in solution, 1 mL of the [209]Po tracer was added. This was not the case with samples #1 and #2, where the [209]Po tracer was added to solution prior to dissolution. All solutions were transferred from the dissolution vessels to 50-mL centrifuge tubes. Each solution was very concentrated and a white crystalline precipitate had formed in large quantities in all of the lead containing samples. The precipitate was dissolved via dilution with deionized water. Three of the sample sets (#7, #8, and #10) contained a secondary precipitate that was a very fine, gray powder. This secondary precipitate would not re-dissolve under the conditions present, and was discarded along with the $PbCl_2$ formed in the following step.

Concentrated HCl was added to each solution until the formation of additional precipitate was no longer observed. The precipitate ($PbCl_2$) was a fluffy, easily suspended, white material. A centrifuge was used to remove it from the bulk solution. The centrifuge was set to run at 2000 rpm for 10-minute intervals until the precipitate formed a puck and no longer suspended in the solution upon gentle swirling of the centrifuge tube. The supernate was poured into a 150-mL beaker and evaporated to near-dryness using a standard hot plate set no higher than 185°C. The precipitate was rinsed with 3 M HCl, mixed with a pressure-activated GE sonicator, and then the resulting solution was added to the beaker containing the supernate. The precipitate was discarded. The supernate was transposed into 5 mL 3 M HCl before separating Pb, Bi, and Po using Eichrom® Sr Resin cartridges (50-100 µm, Lot SRSR11D) and a vacuum box.

Each cartridge was preconditioned with 5 mL 3 M HCl. Each sample aliquot was loaded onto a cartridge and the 150-mL beakers were quantitatively rinsed with 10 mL of 3 M HCl (3 x 3.33 mL), which was also added to the cartridge. This load solution and an additional 20 mL of 3 M HCl were collected as the Pb fraction. Bismuth was eluted in 20 mL of concentrated HCl. Finally, Po was eluted in 40 mL 8 M $HNO_3$ and transferred into a 50-mL glass beaker. These solutions were then transposed into 3 M HCl. The resulting solutions were deposited onto silver discs via spontaneous deposition in a heated hydrochloric matrix (100°C in 3M HCl) for 2 hours. These discs were then analyzed via α spectrometry. Commercially-available Canberra PIPS detectors were used to acquire seven days of counting data on each sample aliquot. Raw data are recorded as α spectra and count rates are determined for the 4.976 MeV [209]Po (half-life: 125 years; α-branch: 99.52%) and 5.407 MeV [210]Po (half-life: 138.4 days; α-branch: 100%) α particles from each isotope.

## Measurement of [210]Bi bremsstrahlung in Lead using γ-spectrometry

The presence of [210]Bi (half-life: 5.012 days; β-branch: ~100%) in lead as a daughter isotope of [210]Pb induces bremsstrahlung radiation when the [210]Bi nuclei beta-decay electron (1.162 MeV beta-decay end point energy) loses energy through scattering within the lead bulk [7,8]. The measurement method presented here employs a low-background, 14-crystal HPGe γ-ray spectrometer [9] operating in the PNNL shallow underground laboratory [10] to measure the [210]Bi-decay-induced bremsstrahlung spectrum. Five of the lead samples assayed via α spectroscopy were selected, spanning the range of ~1 – 75 Bq/kg [210]Pb content, as inferred from the [210]Po α-based assay method as described in the results and discussion section of this article. Lead remaining from the original five bricks used in the [210]Po α-based assay was machined into a standard geometry of 1 cm × 10 cm × 10 cm, as shown in Figure 1. The prepared lead samples were washed with a 2% solution of Micro90 (International Products Corp., Burlington, NJ, USA) and rinsed thoroughly with high purity water (>18.2 MOhm), etched for 3 minutes in a 1% (v/v) nitric acid and 3% hydrogen peroxide solution, rinsed in high purity water, then rinsed briefly in ethanol and allowed to dry.

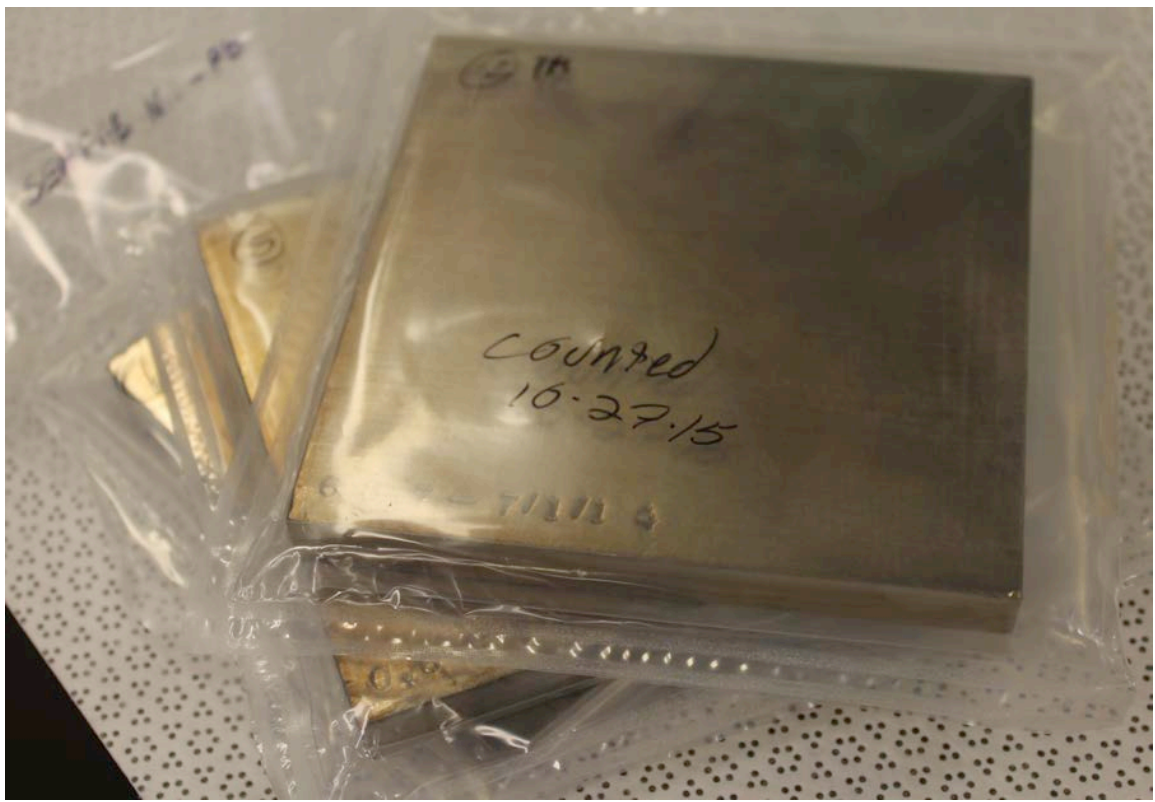

Figure 1. Lead samples prepared in a standard geometry for calibrating the equivalent $^{210}$Pb concentration based on a γ-ray measurement of $^{210}$Bi bremsstrahlung spectrum.

A majority of the $^{210}$Bi bremsstrahlung spectrum observed in the γ-ray spectra appears above the 66.7 keV $^{73m}$Ge line caused by absorption of thermalized neutrons in germanium, and below the 511 keV annihilation peak. K-shell X-ray peaks from lead in the energy range 75 − 85 keV appear in the lead sample spectra but are absent from the system background spectrum. The atomic relaxation process producing these X-rays from the lead samples should scale linearly with the $^{210}$Pb concentration in the same manner as the $^{210}$Bi bremsstrahlung spectrum. For these reasons, the count rate in a region of interest (ROI) spanning 70 − 500 keV was analyzed to establish a calibration between the bremsstrahlung (and X-ray) count rate and the $^{210}$Pb content of the samples determined from the $^{210}$Po α-based assay. To further improve the ROI analysis, four system background peaks were eliminated from the ROI integral count rate in the 70 − 500 keV range by simply ignoring the count rate contribution in 6 keV wide "rejection windows" around the following system background peaks:

1. $^{75m}$Ge at 139.7 keV
2. $^{71m2}$Ge at 198.4 keV
3. $^{212}$Pb at 238.6 keV
4. $^{214}$Pb at 295.2 keV
5. $^{214}$Pb at 351.9 keV

The five prepared lead samples were measured on the 14-crystal germanium array for counting periods ranging from ~1 − 2 days for the higher $^{210}$Pb content samples to ~7 days for the two lowest activity samples. Energy calibration is accomplished independently for each crystal in the array, and the native 65536 channel spectral data from the XIA DGF4 digitizers are then rebinned into a common 8192-channel spectrum structure. Ignoring the count rate contributions in the four rejection windows described above, data for the channels spanning 70 to 500 keV were summed to determine a single gross count rate for the ROI. The one-sigma uncertainty in the gross count rate was calculated as the square root of the total counts collected in the ROI, divided by the count duration. Figure 2 shows the spectrum (binned into fewer channels to improve clarity of the figure) from three of the lead standards, as well as the typical system background. It is interesting to note that the continuum from the lowest activity standard is below that of the system background; this is reasonable, because this lead standard is significantly lower in $^{210}$Pb activity than the system shield, and provides some additional shielding between the individual detector crystals.

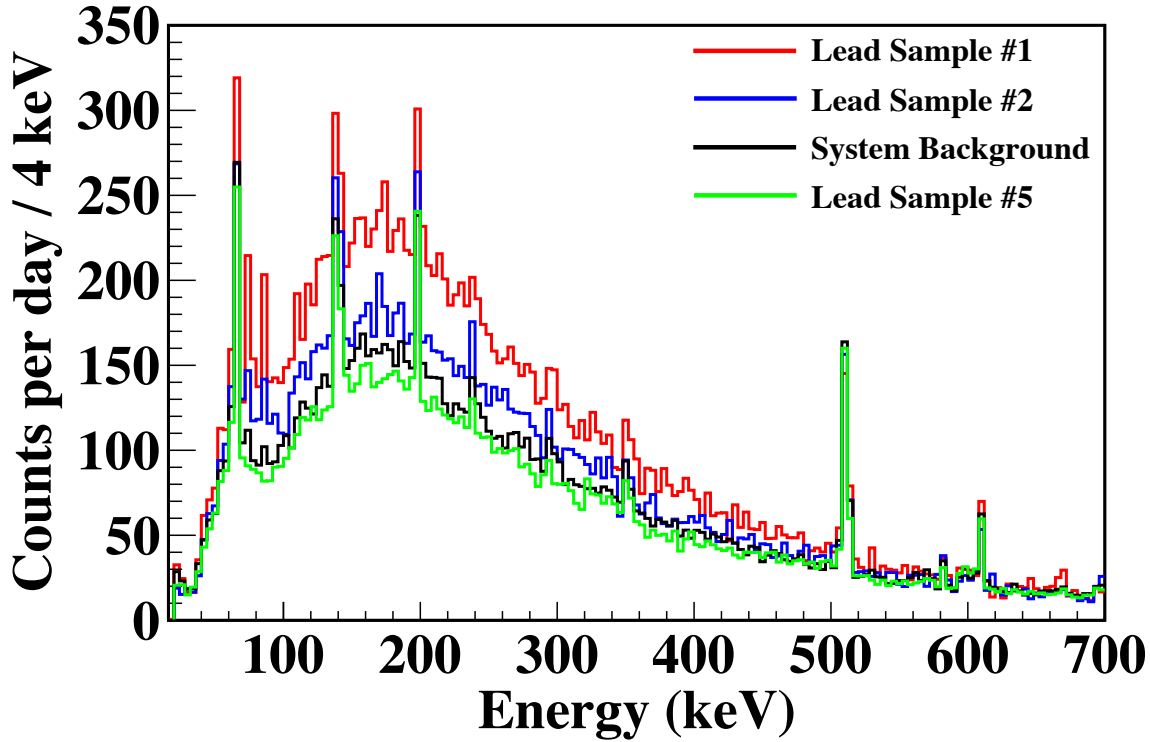

Figure 2. Comparison of the spectra from three lead samples and the system background.

### Lead Sample History

To augment forthcoming discussion of analytical results from the measurements described above, short descriptions of the histories of the ten lead samples are given. These histories are based on word of mouth and personal recollections. The reader is therefore cautioned from drawing strong conclusions regarding the relationship between the analytical results presented and the specific histories described.

*#1 – PNNL ~50 year old stock.* This lead was used in room 17-A of the now demolished 329 building at Pacific Northwest National Laboratory (PNNL) in Richland, WA. For many decades this room served as a dedicated, low-background counting room. The lead was used as a large, single brick-thick, floor-to-ceiling, multi-alcove detector shielding arrangement (See Fig. 1 in both [11] and [12]). The lead was likely purchased, with no special specifications, in the middle of the 1960's although the Hanford 300 area buildings were originally constructed in the early 1950's. Much of the lead was re-purposed for use in the shield of the 14-crystal HPGe array detector [1] used in this report.

*#2 – Doe Run (Sullivan vendor).* This lead was purchased from Sullivan Metals, Inc. of Holly Springs, MS in 2010. The lead is originally from the Doe Run Company, mined from the Viburnum Trend in Southeast Missouri. The bricks are used in the Radionuclide Laboratory 16 (RL16) HPGe detector shields in the shallow underground laboratory [2] located at PNNL in Richland, WA. The RL16 detectors support the work of the Preparatory Commission for the Comprehensive Nuclear-Test-Ban Treaty Organization (CTBTO) [13,14].

*#3 – University of Chicago stock.* This lead is in use by Dr. Juan I. Collar (University of Chicago) as part of the shield for the C-4 dark matter experiment [15] located at the Soudan Underground Laboratory. The specific history for this lead was not recalled as the lead was not expected (or intended) to be particularly low background.

*#4 – Doe Run (Seafab vendor).* This lead is from the same Doe Run Company source as the lead described for Sample #2. However, this lead was purchased in 2014 from Seafab Metals Company of Casa Grande, AZ. This lead was used in the construction of the middle layer of the shield for a low background liquid scintillation counter [16] located in the PNNL shallow underground laboratory [2] in Richland, WA.

*#5 – PNNL ancient lead.* This lead is Spanish galleon ballast lead that was re-smelted at PNNL and cast into bricks in the 1980's. Upon machining, it was found these bricks had voids within the material. The voids were not large enough or prevalent enough to cause a noticeable difference in total brick weight, however this observation shows these bricks are potentially non-homogenous. This lead was used in the construction of the inner layer of the shield for a low background liquid scintillation counter [16] located in the PNNL shallow underground laboratory [2] in Richland, WA.

*#6 – LANL stock.* This lead is at least 150 years old and started as sheets in the Boston sewer system. At some point in the 1960's the lead was cast into bricks and used by the Air Force at McClellan AFB near Sacramento, CA as shielding in a low background counting facility. In approximately 1999, with the start of decommissioning McClellan AFB, the bricks were shipped to Los Alamos National Laboratory (LANL) for use as shielding in a counting facility. Much of the lead was stored on an asphalt pad at TA-33 near Bunker 22 for approximately 15 years as the property of C Division[1].

*#7 – University of Chicago Spanish lead.* This lead is also antiquity, Spanish galleon ballast lead. This lead is in use by Dr. Juan I. Collar (University of Chicago) as part of the shield for the C-4 dark matter experiment [15] located at the Soudan Underground Laboratory.

*#8 – Hampton Court Palace roof.* This lead was cast into ingots stamped with the title "AMALGAM" and is lead recast from the lead shingles of the Hampton Court Palace roof. This same source of lead was used in another low background HPGe detector system and was stated to have 20 Bq/kg of $^{210}$Pb in Hult *et al*. [4]. However, following the references to an article by Mouchel and Wordel [17], it is seen the "(d) English monument" lead[2] listed in Table 3 is a factor of 9 higher than the ≤2 Bq/kg level observed in the "(f) French monument" lead. In Mouchel and Wordel [17], only the single "(f) French monument" lead sample is given a quantitative value for $^{210}$Pb concentration, and that value is an *upper limit*. The results from this work suggest the actual concentration in the Hampton Court Palace roof lead is ~1 Bq/kg.

*#9 – PNNL Sequim stock.* This lead was used in a low background shield at the PNNL Marine Sciences Laboratory in Sequim, WA. It was included in this analysis to determine if it was particularly low background and therefore to be kept for future use.

*#10 – PNNL German lead.* This lead brick was from the inner, low background lead liner of the IGEX neutrinoless double beta decay experiment when it was located at the Homestake mine, in Lead, SD [18]. It is the same German lead used in the SOLO detector [19] at the Soudan Underground Laboratory.

---

[1] Thank you to K. Rielage (Los Alamos National Laboratory) for the description of the sample provided.
[2] The Acknowledgements in Mouchel and Wordel [17], state an old lead sample – presumably the "(d)
[2] The Acknowledgements in Mouchel and Wordel [17], state an old lead sample – presumably the "(d) English monument" of Table 3 – is "from the roof of Hampton Court".

## Results and discussion

Results from the determination of $^{238}U$ and $^{232}Th$ in lead using ICP-MS are presented in Table 1. The measurement of the $^{238}U$ concentration was limited in this analysis by the size of the initial lead aliquot samples provided for the $^{238}U$ and $^{232}Th$ assay. Additional work has shown sensitivity to $^{238}U$ concentration in lead matrix can reach the level of ~1 microBq/kg if sufficient lead matrix is digested and processed. There is little notable regarding the concentration levels of $^{232}Th$ measured in the ten lead samples, other than the roughly factor of 10 higher level seen in the Hampton Court Palace roof lead. Recall this lead sample (#8) showed a significant amount of indeterminate precipitate during the preparation of the $^{210}Po$ α-spectroscopy analysis (see Figure 3).

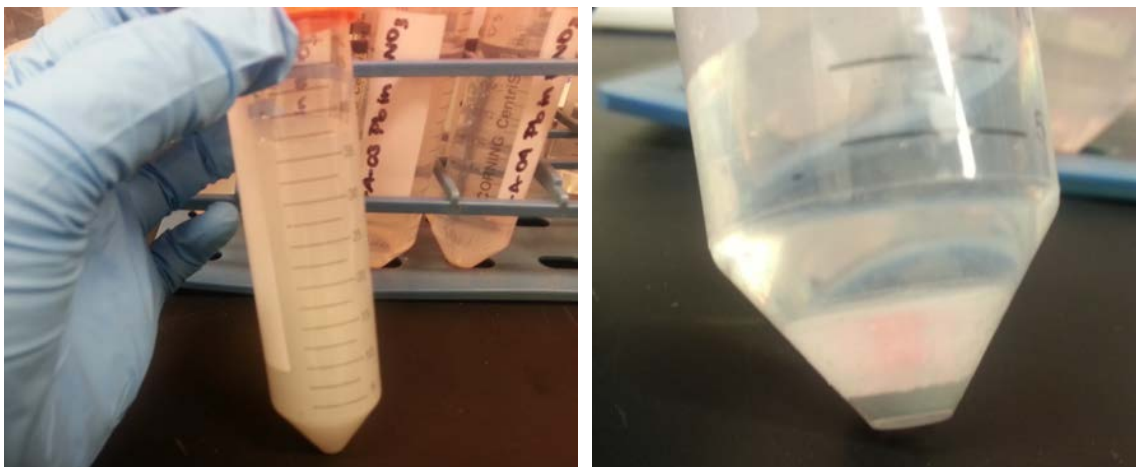

Figure 3. Photographs of the unexpected secondary precipitate seen in the Hampton Court Palace roof lead (Sample #8) during the $^{210}Po$ α-spectroscopy assay preparation. The precipitate would remain suspended for several hours before settling to the bottom on the vials. The nature of the precipitate was not investigated.

To report the results from the $^{210}Po$ α-spectroscopy analysis in Bq $^{210}Pb$/kg, a calculation is performed under the assumption that all ten lead samples are sufficiently old enough to ensure secular equilibrium has been established between the $^{210}Pb$ parent and the daughter isotopes of $^{210}Bi$ (half-life: 5 days) and $^{210}Po$ (half-life: 138 days). Based on the lead histories described above, this assumption appears reasonable. It is worth noting that freshly refined commercially available lead may not satisfy this secular equilibrium requirement on the several month time-scale. This can result in a discrepancy between $^{210}$Pb concentration levels inferred (separately) from $^{210}$Bi bremsstrahlung γ-spectroscopy and $^{210}$Po α-spectroscopy measured during the several month time period after refinement. In such cases, the first daughter $^{210}$Bi, having a short 5-day half-life, provides the more appropriate method for quantification. This is one rationale for the development of the $^{210}$Bi bremsstrahlung γ-spectroscopy measurement described below.

Returning to the reporting of the $^{210}$Po α-spectroscopy analysis, the calculation to infer the $^{210}$Pb concentration in units of Bq/kg proceeds as follows. For each sample's four aliquots, the seven-day α-spectroscopy spectrum is analyzed to determine the gross count rates in peaks associated with the $^{209}$Po 4.976 MeV α particle and $^{210}$Po 5.407 MeV α particle. The equation used for each lead sample aliquot is

$$\left[ ^{210}\text{Pb (Bq/kg)} \right] = \frac{\left( \left( R_{210}^{\text{peak}} \middle/ \left( R_{209}^{\text{peak}} \middle/ R_{209}^{\text{tracer}} \right) \right) - R_{210}^{\text{blank}} \right)}{M_{\text{Pb}}}$$

where the concentration of $^{210}$Pb is reported in Bq/kg based on the α-particle rate in the $^{210}$Po 5.407 MeV peak ($R_{210}^{\text{peak}}$) divided by the $^{209}$Po tracer recovery efficiency determined from the α-particle rate in the $^{209}$Po 4.976 MeV peak ($R_{209}^{\text{peak}}$) divided by the known tracer amount in Bq ($R_{209}^{\text{tracer}}$), with the process blank α-particle rate in the $^{210}$Po 5.407 MeV peak ($R_{210}^{\text{blank}}$) subtracted from the result. The mass of the lead sample aliquot, $M_{\text{Pb}}$, is recorded in kilograms. The 1-sigma errors were propagated through the equation. To generate a combined result for each lead sample, the four individual aliquot results were combined in a simple, error-weighted average. In some of the lead sample cases, not all of the four aliquot measurements were successful; some were lost in preparation. The simple, error-weighted average $^{210}$Pb concentration inferred from the $^{210}$Po α-spectroscopy assay are presented in Table 1. The lead samples having an asterisk (*) had an issue with the $^{210}$Po assay (either a lost aliquot or only upper limits were achieved).

The lead samples having a diamond (◊) were used to create the calibration standards for the γ-spectrometry measurement described in more detail below.

Table 1. Results from the ICP-MS measurement of $^{238}$U and $^{232}$Th and α-spectroscopy measurement of $^{210}$Po. The results from the $^{210}$Po measurement are reported in units of $^{210}$Pb concentration following the calculation procedure described in the text body. See text for further description of the lead samples.

| Lead sample | Sample # | $^{238}$U (microBq/kg) | $^{232}$Th (microBq/kg) | $^{210}$Pb (Bq/kg) |
|---|---|---|---|---|
| PNNL ~50 year old stock ◊ | #1 | < 34 | < 0.7 | 68.7 ± 1.5 |
| Doe Run (Sullivan vendor) ◊ | #2 | < 16 | 1.7 ± 0.1 | 30.2 ± 0.6 |
| U. Chicago stock* ◊ | #3 | < 22 | 0.5 ± 0.2 | 74.3 ± 1.9 |
| Doe Run (Seafab vendor) | #4 | < 23 | 1.9 ± 0.3 | 29.4 ± 0.7 |
| PNNL ancient lead ◊ | #5 | < 30 | 1.6 ± 0.1 | 0.09 ± 0.01 |
| LANL stock* ◊ | #6 | < 30 | < 0.6 | 8.7 ± 0.2 |
| U. Chicago Spanish lead* | #7 | < 23 | 2.2 ± 0.1 | 0.03 ± 0.01 |
| Hampton Court Palace roof | #8 | < 29 | 15 ± 0.6 | 1.34 ± 0.04 |
| PNNL Sequim stock | #9 | < 17 | 1.7 ± 0.2 | 58.5 ± 1.3 |
| PNNL German lead* | #10 | < 33 | 3.0 ± 0.2 | 0.13 ± 0.02 |

Using the results from the $^{210}$Po α-spectroscopy assay presented earlier in this article, a set of calibration standards was developed for determining the $^{210}$Bi concentration in lead based on bremsstrahlung radiation emission levels. The approach described here to calibrate the $^{210}$Pb concentration level in lead using γ-spectrometry is conceptually an experimental validation of the work previously developed by Nachab and Hubert [20]. In that work [3] a calibration method between the measured $^{210}$Bi-decay induced bremsstrahlung spectrum and the supporting $^{210}$Pb activity was accomplished via Monte Carlo simulation modeling. Furthermore, the simulation work of Vojtyla [21] is often used for evaluating the bremsstrahlung emission from a solid lead shield. The work presented in this report empirically supports these simulation-based methods. As the $^{210}$Bi bremsstrahlung γ-ray assay method for evaluating the $^{210}$Pb concentration in lead relies upon the underlying radiochemical separation and α-spectroscopy measurements, attention to the details of the $^{210}$Po α-spectroscopy assay method are scrutinized for their veracity. To this end, Figure 4 presents the individual aliquot measurements calculated for $^{210}$Pb concentration.

In Figure 4, Sample #1 clearly has a well-defined (error weighted) mean value from three of four aliquot measurements. Sample #2 also has a well-defined (error weighted) mean with one of the four aliquot measurements appearing as an outlier. Sample #3 shows variation at roughly the two-sigma level, but can reconstruct a (error weighted) mean through normal error propagation. Sample #5 presents the greatest difficulty in interpretation. Recall from the history description of the PNNL ancient lead (Sample #5), the lead bricks had voids implying a non-homogenous bulk. More confounding, two aliquot results (#1 and #3) have reported errors of >20%. In the majority of other aliquot measurements across the entire ten lead samples, the $^{210}$Po α-spectroscopy analysis for the $^{210}$Pb concentration reported errors of 3.5% – 6% on individual aliquot measurements. However, as the absolute value of the inferred the $^{210}$Pb concentration for Sample #5's aliquots #1 and #3, the large reported errors (>20%) appear surprisingly small on an absolute scale. Thus the Sample #5's aliquots #1 and #3 are deemed suspicious and removed from the error weighted mean for Sample #5. Unfortunately, Sample #5's other two aliquots do not appear to agree within their reported errors. In the end, to handle this difficult situation, it was decided to use the error weighted mean from aliquots #2 and #4 for Sample #5 with an arbitrary inflation of the propagated error by a factor of 10. The impacts of these choices will be investigated further during the development of the calibration fit for the $^{210}$Bi bremsstrahlung γ-spectroscopy measurement. Finally, Sample #6 has two data points (aliquots #1 and #2) that agree, with one outlying data point (aliquot #3). This discussion of the data selection to eliminate outlier measurements in the

$^{210}$Pb concentration levels inferred from α-spectroscopy measurements of $^{210}$Po is presented visually in Figure 4.

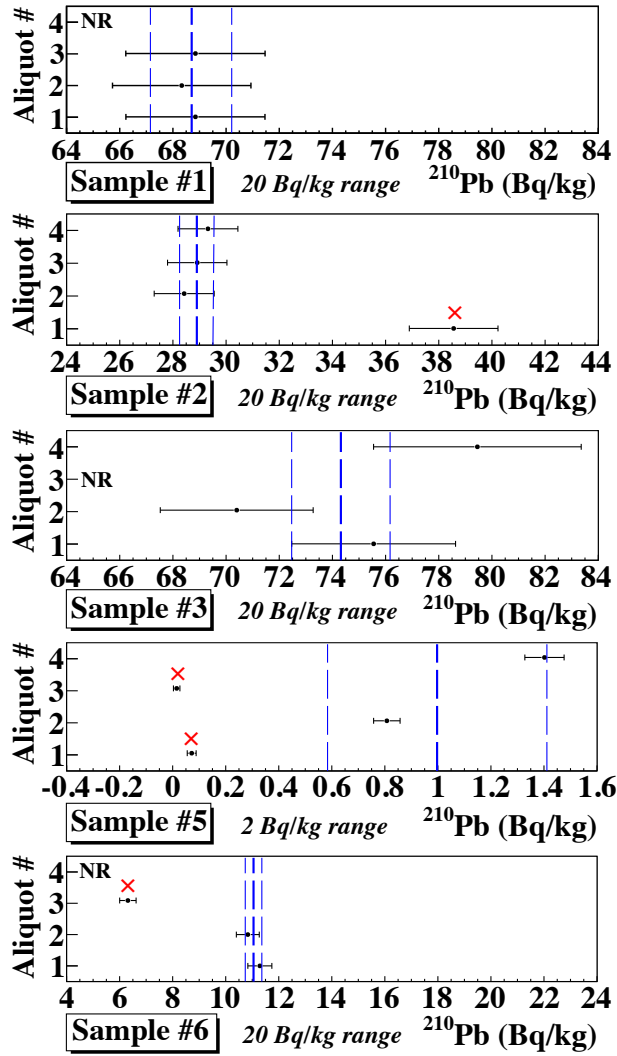

Figure 4. Selection of $^{210}$Pb assay results used to create calibration standards for $^{210}$Pb evaluation via measurement of $^{210}$Bi bremsstrahlung spectra. The abbreviation "NR" is for aliquots lost during chemical processing. A red "×" is placed above outlying data points excluded from the evaluated $^{210}$Pb concentration levels for the five lead samples used as calibration standards. Note the $^{210}$Pb concentration range (x-axis scale) for Sample #5 is a factor of 1/10 of the range presented for the other four samples.

From the foregoing discussion of removal of outliers in the $^{210}$Po α-spectroscopy analysis for $^{210}$Pb concentration levels, Table 2 presents the final data used to create a calibration of $^{210}$Pb concentration based on the gross count rate in a region of interest (ROI) from the $^{210}$Bi bremsstrahlung γ-spectroscopy measurement results. For ease of comparison the simple, error-weighted $^{210}$Pb concentration levels derived from the $^{210}$Po α-spectroscopy analysis originally presented in Table 1 are shown again in Table 2.

Table 2.    Analysis results used in the development of the calibration of the $^{210}$Bi bremsstrahlung γ-spectroscopy measurement method for determination of $^{210}$Pb concentration levels in bulk refined lead samples.

| Lead sample | Sample # | $^{210}$Pb *Simple Weighted* (Bq/kg) | $^{210}$Pb *Outliers Removed* (Bq/kg) | Gross Count Rate in ROI (cpd) |
|---|---|---|---|---|
| PNNL ~50 year old stock | #1 | 68.7 ± 1.5 | 68.7 ± 1.5 | 13275 ± 85 |
| Doe Run (Sullivan vendor) | #2 | 30.2 ± 0.6 | 28.9 ± 0.6 | 10482 ± 70 |
| U. Chicago stock | #3 | 74.3 ± 1.9 | 74.3 ± 1.9 | 13859 ± 118 |
| PNNL ancient lead | #5 | 0.09 ± 0.01 | 1.00 ± (0.04×10) | 8300 ± 35 |
| LANL stock | #6 | 8.7 ± 0.2 | 11.0 ± 0.3 | 9068 ± 37 |

Figure 5 presents results from a Deming linear regression [22] (as implemented in SigmaPlot for Windows, Build 12.5.0.38) of the HPGe measurement of bremsstrahlung continuum vs. the $^{210}$Pb concentration based on $^{210}$Po α-spectroscopy assay. Results of the fit are a y-intercept of 8240(35) cpd, and a slope of 75.2(1.6) (cpd)/(Bq/kg). The y-intercept represents the system background continuum rate, with a notional $^{210}$Pb-free lead piece in place, and contains contributions from a combination of bremsstrahlung from the lead shield, residual un-vetoed cosmic radiation, and a small number of Compton scattering events.

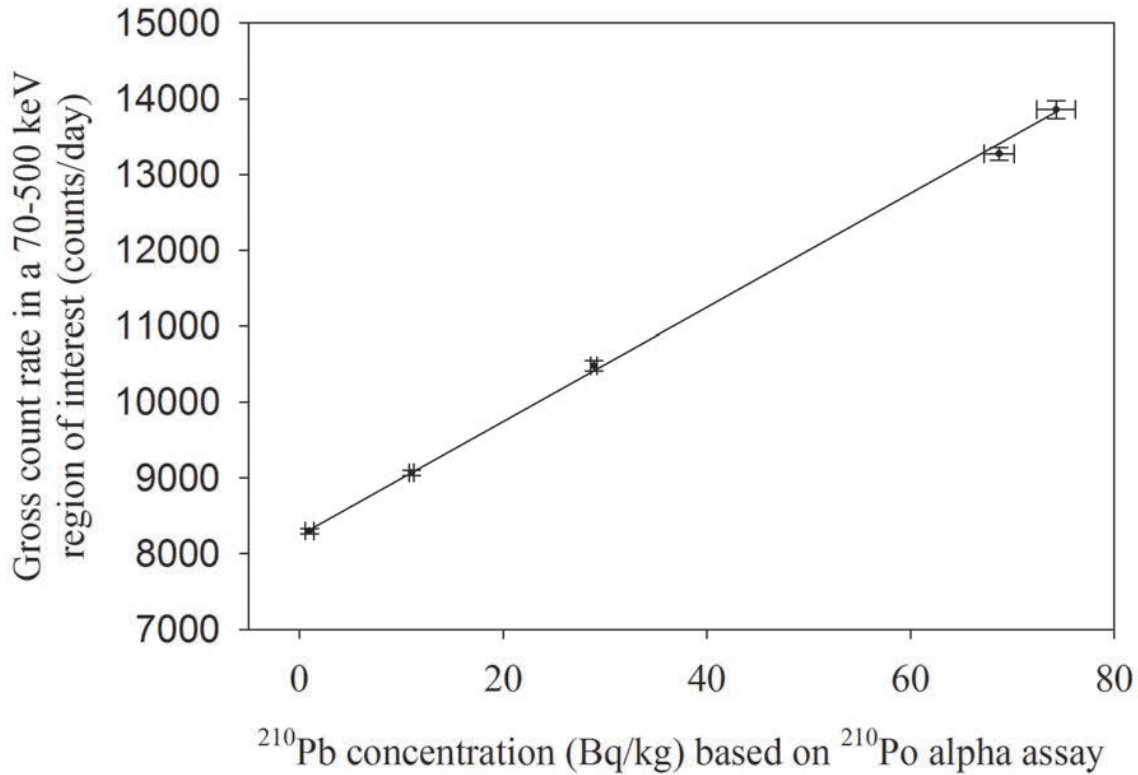

Figure 5.  Deming linear regression for evaluating [210]Pb concentration via bremsstrahlung γ-assay based on [210]Po α-spectroscopy assay measurements for five lead samples.

The impact of the outlier data rejection choices discussed above and in Fig. 4 were investigated by considering the linear regression outcomes using alternative selections, including no data points rejected, not increasing the uncertainty of Sample #5, and setting Sample #5 to have no [210]Pb content (0.0±0.01Bq/kg). Above 10 Bq/kg, these various cases had little impact on the result calculated for an unknown sample assay. The discrepancy is more significant for assay results between 2 and 10 Bq/kg, however consistent with the uncertainties that would be quoted with the results described below.

Figure 6 provides a family of curves that show the count length required to achieve a desired uncertainty. This chart shows that a [210]Pb content of ~12 Bq/kg or higher can be determined to 5% or better uncertainty via the γ-ray assay method. However, for low background lead of ~2 Bq/kg $^{210}$Pb content, only a 25% uncertainty measurement is expected via the γ-ray assay method. Improved knowledge of the γ-ray background continuum count rate could make a small improvement in the lowest concentration of lead that can be assayed. However, further reduction of the system background continuum is necessary to make significant gains. That said, the available range for assay of $^{210}$Pb concentration is useful for most modern refined lead, and many typical lead samples would only require a day or two of measurement for better than 5% uncertainty. Assay of old lead stocks reaching down to 0.1 Bq/kg or better would be possible on deep underground γ-ray spectrometers with lower continuum (e.g., the GeMPI detectors at Gran Sasso [23]).

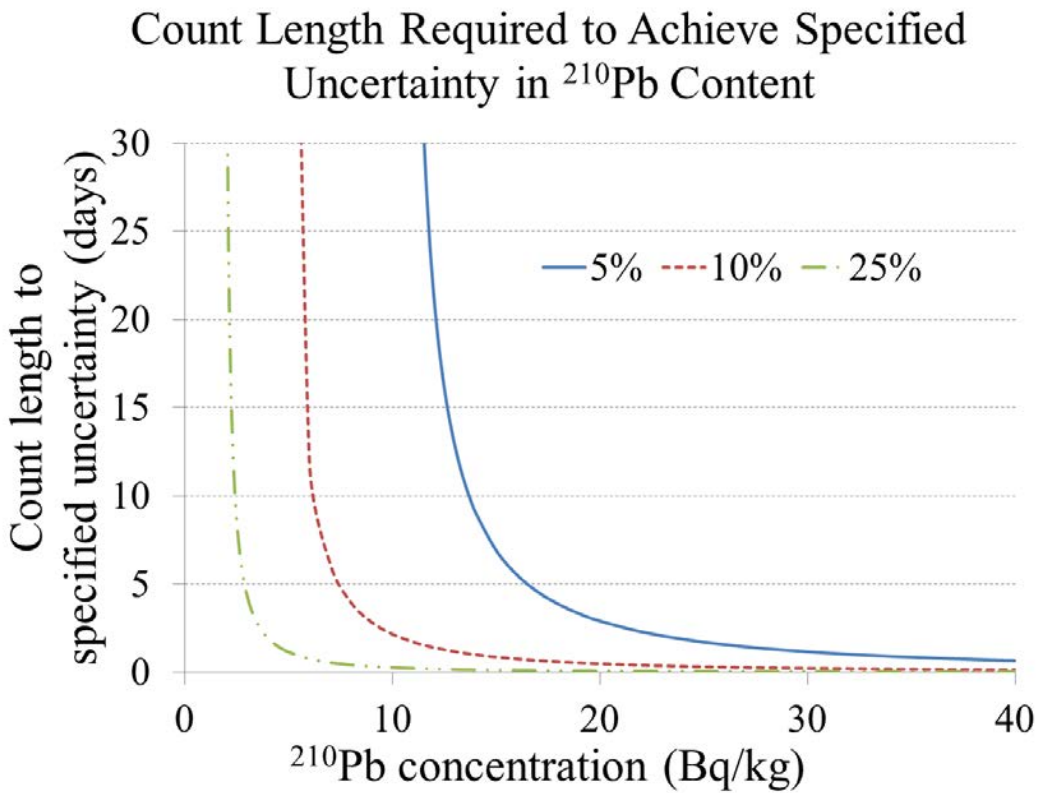

Figure 6. A family of curves showing the count length required for achieving a desired uncertainty in the $^{210}$Pb concentration assay based on $^{210}$Bi bremsstrahlung emission.

## Conclusions

The continued development of low background radiation detection systems for research and applications in physics and environmental monitoring will require the ability to effectively and efficiently screen shielding materials for radioactive content to ensure the background targets of the designed systems are achieved. Commonly used as a γ-ray shielding material, lead typically requires evaluation of the $^{238}$U, $^{232}$Th, and $^{210}$Pb concentrations to determine the appropriateness for a given low background shielding design. This article reports methods for assay of all three radioactive isotopes and demonstrates these assay methods on ten lead samples of various origins. Additionally, five $^{210}$Pb standards were created and used to calibrate a 14-crystal HPGe array for $^{210}$Bi bremsstrahlung γ-ray emission as a function of $^{210}$Pb concentration in a solid lead 1 cm × 10 cm × 10 cm geometry. Researchers interested in borrowing the five solid lead 1 cm × 10 cm × 10 cm geometry $^{210}$Pb-concentration standards for system calibration purposes are encouraged to contact the authors of this article.

## Acknowledgements

The three radiochemical assay measurements for $^{238}$U, $^{232}$Th, and $^{210}$Po described in this article were supported by the Ultra Sensitive Nuclear Measurements Initiative, conducted under the Laboratory Directed Research and Development (LDRD) Program at Pacific Northwest National Laboratory, a multiprogram national laboratory operated by Battelle for the U.S. Department of Energy.

with the sensitivity of ββ-decay spectrometers. Nuclear Physics B - Proceedings Supplements 143:564. doi:10.1016/j.nuclphysbps.2005.01.228